\documentclass{article}

\usepackage[english]{babel}

\usepackage[letterpaper,top=2cm,bottom=2cm,left=3cm,right=3cm,marginparwidth=1.75cm]{geometry}

\usepackage{amsmath}
\usepackage{graphicx}
\usepackage{authblk}
\usepackage[colorlinks=true, allcolors=blue]{hyperref}

\usepackage[authoryear,round]{natbib}

\newcommand{\hi} {H {\sc i}}
\newcommand{\lalpha}{Lyman-$\alpha$}
\newcommand{\cii}{C {\sc ii}}
\newcommand{\hei} {He {\sc i}}
\newcommand{\pimenc}{$\pi$ Men c}
\newcommand{\hep}{He$^+$}
\newcommand{\heii} {He {\sc ii}}
\newcommand{\eminus} {e$^-$}
\newcommand{\htwo}{H$_2$}

\title{Star-Planet Interactions: Observational Techniques and Methods}

\author[1,2]{P.~Figueira\thanks{corresponding author; \texttt{pedro.figueira@iaa.es}}}
\author[3]{H.~Korhonen}
\author[4,5]{A.~Buccino}
\author[6]{P.~Chaturvedi}
\author[7]{R.~Fares}
\author[8]{A.~Garc\'ia-Mu\~noz}
\author[9]{B.~Montet}
\author[2]{L.~Pe\~na-Mo\~nino}
\author[2]{M.~P\'erez-Torres}
\author[2]{D.~Revilla}
\author[10]{A.~Valio}

\affil[1]{Observatoire Astronomique de l’Universit\'{e} de Gen\`{e}ve, Chemin Pegasi 51b, 1290 Versoix, Switzerland}
\affil[2]{Instituto de Astrof\'{i}sica de Andaluc\'{i}a-CSIC, Glorieta de la Astronom\'{i}a s/n, 18008 Granada, Spain}
\affil[3]{Max-Planck-Institut für Astronomie, Königstuhl 17, D-69117 Heidelberg, Germany}
\affil[4]{Departamento de Física, FCEyN Universidad de Buenos Aires, Buenos Aires, Argentina}
\affil[5]{Instituto de Astronomía y Física del Espacio (UBA-CONICET), Buenos Aires, Argentina}
\affil[6]{Department of Astronomy and Astrophysics, Tata Institute of Fundamental Research, Mumbai, India}
\affil[7]{Department of Physics, College of Science, United Arab Emirates University, P.O. Box 15551, Al Ain, UAE}
\affil[8]{Universit\'e Paris-Saclay, Universit\'e Paris Cit\'e, CEA, CNRS, AIM, 91191, Gif-sur-Yvette, France}
\affil[9]{School of Physics, University of New South Wales, Kensington 2052, Australia}
\affil[10]{Centro de Rádio Astronomia e Astrofísica Mackenzie (CRAAM), Universidade Presbiteriana Mackenzie, São Paulo, Brazil}

\date{}

\begin{document}
\maketitle

\begin{abstract}
\noindent This chapter summarizes the techniques and methods used to study star-planet interaction (SPI) from the observational point of view. 
SPI can produce a wide range of observational signatures, from localized stellar activity enhancements to changes in planetary atmospheric escape and transmission spectra. This chapter reviews the main observational techniques used to detect and characterize SPI, emphasizing the methodological challenges involved in separating planet-induced signals from intrinsic stellar variability. We discuss radial-velocity diagnostics, including cross-correlation, template matching, line-by-line methods, and activity indicators, highlighting their sensitivity to line-profile distortions and chromatic variability. We then review precision photometry as a tool to search for orbit-locked variability, flare modulation, and active-region occultations during planetary transits, with particular attention to detrending, correlated-noise modeling, Gaussian processes, and statistical validation. Chromospheric diagnostics, including Ca\,{\scriptsize{II}}, H$\alpha$, He\,{\scriptsize{I}}, and Na\,{\scriptsize{II}} lines, are presented as tracers of magnetic variability at different atmospheric heights and as potential probes of intermittent SPI signatures. We also discuss transmission spectroscopy as a complementary approach, since planetary atmospheric tracers such as H\,{\scriptsize{I}} Ly$\alpha$, Balmer lines, C\,{\scriptsize{II}}, and He\,{\scriptsize{I}} can encode information about the stellar high-energy environment, stellar wind, and magnetic coupling. In addition, radio observations provide a promising avenue to probe magnetic SPI directly through coherent emission mechanisms, offering unique constraints on planetary magnetic fields and star–planet coupling. Finally, we examine time-series analysis techniques commonly employed in SPI searches, including generalized Lomb--Scargle periodograms, rolling periodograms, harmonic analyses, and bootstrap-based significance estimation, with emphasis on the identification of transient and orbit-related signals in unevenly sampled datasets. Together, these methods show that robust SPI identification requires coherent evidence across timescales, diagnostics, and atmospheric layers, rather than a single periodic signal.\footnote{The instrumentation that employs these methods is presented in the companion chapter Korhonen et al. (2026).} \\
  
\end{abstract}

\textit{The authors have no relevant financial or non-financial interests to disclose, and no competing interests to declare that are relevant to the content of this article.}

\textit{The authors have no conflicts of interest to declare.}

\section*{Introduction}

The methods used to study SPI are as diverse as the range of possible star–planet interaction signatures. Moreover, the difficulty of establishing a “smoking gun” signature—one that unambiguously confirms the interaction—has led to a proliferation of detection and characterization efforts. Here, we outline the various methods employed in the study of star–planet interactions (SPI), either by explaining their fundamental principles in the context of SPI or by introducing them in general terms and subsequently discussing their specific applications to SPI.

\vspace{2.0cm}

\section*{Radial Velocities}


\noindent Precise radial velocity (RV) measurements have been fundamental to our understanding of stellar properties in the solar neighbourhood \citep{DuquennoyMayor91}. Following the detection of the first planet orbiting a Sun-like star by \citet{MayorQueloz95}, this technique came under close scrutiny. Because of its sensitivity to variations in the positions and shapes of photospheric lines, it also traces stellar activity and, by extension, activity induced by the presence of a planet.

We group the different RV calculation methods according to their underlying assumptions and methodologies, and complement these by a brief description of associated activity indicators. We conclude the section by discussing how these RV signatures can be identified in time series data in practice.

\subsection*{Fourier domain methods and spectral information content}

A foundational approach to RV calculation was outlined by \citet{Cheli20}, who considered the case of an observed spectrum that differs from a reference spectrum only through an RV shift and the presence of photon noise. The author provided a framework for an optimally weighted RV calculation in Fourier space. In practice, the reference spectrum can be constructed by co-adding a large number of spectra, and working in Fourier space allows the filtering of both high and low frequencies, corresponding to photon noise and continuum variations, respectively. For an efficient application of this framework to A–F main-sequence stars, the reader is referred to \citet{Galland05}.

The optimal weighting principles proposed by \citet{Cheli20} define thus a best-case scenario for RV calculation, and motivated the study of spectral information content by \citet{Bouchy01}. These authors provided a methodology, this time in the wavelength domain, both to calculate the RV and to estimate the RV precision achievable for a given spectrum in the presence of photon noise alone. This method has since become invaluable for estimating the precision achievable in the high-S/N regime, as is often reached in RV surveys of bright stars. Moreover, it provides an important physical insight: the RV information content of a spectral line is proportional to the local slope of the flux and is therefore concentrated in the \textit{wings} of the line.

While this method was a cornerstone in the development of precise RV measurements, it assumes that the spectrum remains unchanged apart from an RV shift and noise. As such, it cannot identify signatures that depend on wavelength or spectral lines, nor those that alter the global properties of the spectrum. Its usage in SPI studies is therefore limited.

\subsection*{The cross-correlation function}

The cross-correlation function (CCF) method, used in the detection of the first extrasolar planets around main-sequence stars, makes fewer assumptions than Fourier-domain approaches. In a nutshell, an observed spectrum is cross-correlated with a “mask”, i.e. a template containing the rest wavelengths of selected spectral lines. The resulting cross-correlation function can be interpreted as an average line profile in velocity space, from which RVs are typically measured with a simple Gaussian fitting. The method is presented in detail by \citet{Baranne96}. The work of \citet{Pepe02} further developed this approach by introducing weighted lines, i.e., assigning each line a weight proportional to its information content. In early implementations, these weights were simply proportional to the line contrast; in modern pipelines, they are computed from detailed line properties and instrumental characteristics \citep[e.g.][]{Figueira18}.

When considering how to optimally combine cross-correlation functions, it is particularly insightful to revisit \citet{Zucker03}, who provided a statistically rigorous framework for their combination. However, this method draws its strength and rigour by remaining fundamentally tied to the assumption that the observed spectrum differs from the template primarily through a Doppler shift. This undermines its diagnostic power for SPI signatures that manifest as line-dependent or chromatic spectral distortions, a limitation  common to all similar approaches.

Within the context of CCF analysis, \citet{CollierCameron21} represents a significant step forward. This work introduced a wavelength-domain, data-driven method to separate genuine planetary reflex Doppler shifts from apparent RV variations induced by stellar variability. The method exploits the fact that the autocorrelation of the CCF is invariant to RV shifts, whereas stellar activity induces distortions in line profiles. It then uses principal component analysis of the CCF autocorrelation to isolate these shape-driven modes. Because SPI-related signals may manifest as line-profile or wavelength-dependent variability rather than as pure center-of-mass motion, methods such as that of \citet{CollierCameron21} are particularly valuable for distinguishing physically distinct sources of apparent RV variability. This method, named SCALPELS, has been successfully applied to very high-S/N solar spectra and holds significant promise for studying SPI in bright stars.

\subsection*{The template matching approach}

As exoplanet searches began to target M dwarfs, it became apparent that applying the CCF method to these stars was considerably more challenging. Constructing line masks was difficult, as theoretical spectra suffered from incomplete line lists and the numerous but shallow lines of M dwarfs remained poorly defined even in observed spectra with high S/N. In addition, their spectra are strongly affected by pseudo-continuum absorption, and the stars themselves are relatively faint for most spectrographs. These three difficulties  worked together quite well, much to the frustration of observers. 

\citet{Kurster03} proposed a different approach: calculating RVs through a $\chi^2$- match between the observed spectrum and a shifted template spectrum. High-S/N templates could be constructed by co-adding many observed spectra, and the minimum of $\chi^2$(RV) yielded the stellar RV. This allowed to reach a precision of 2.7\,m/s on Barnard’s star using UVES, an impressive result for the time. It laid part of the methodological foundation for the landmark detection of Proxima b by \citet{Anglada-Escude16} using HARPS spectra. In that work, the authors revealed the presence of a short-period planet around our nearest stellar neighbour, showcasing the power of template matching.

Today, several template-matching methods exist and are routinely used. \texttt{SERVAL} \citep{Zechmeister18}, widely used with CARMENES \citep[e.g.][]{Trifonov2018}, is probably the best known. \texttt{S-BART} \citep{Silva22} provides a more modern implementation with somewhat different technical choices. It is worth noting that, to date, one of the most precise long-term RV baselines has been obtained with \texttt{S-BART} on ESPRESSO data: 40\,cm/s over 3.5\,yr on $\tau$ Ceti \citep{Figueira25}. The modest gain in precision for this K dwarf relative to the CCF method may result from a better leveraging of the spectral information content, including the continuum, or from a greater robustness to flux systematics, although  it is impossible to be sure and the method is still under active investigation. \citet{Silva25} recently highlighted that template-matching methods can be affected by intra-night RV slopes under specific conditions, and caution is therefore warranted.

Because template matching assumes a constant reference spectrum, it is intrinsically ill-suited to SPI signatures that manifest through changes in the properties of individual spectral lines. Moreover, recent developments have shown that, despite its remarkable precision, the method still involves subtle technical choices with an impact that is not yet fully understood.

\subsection*{Line-by-line methods}

It has long been known that, in the Sun, different photospheric spectral lines exhibit varying sensitivity to the level of stellar activity \citep{Graybook}. This motivated several studies aimed at measuring the RVs of individual spectral lines \citep[e.g.][]{Dumusque18,Cretignier20}. These studies generally rely on a large number of high-S/N spectra, acquired with instruments with highly stabilized instrumental profiles, such as HARPS, to measure the RV variations of individual lines and assess their sensitivity to stellar activity. 

Among the different implementations of the line-by-line approach, \citet{Artigau22} provided a particularly clear formalization of the method, emphasizing the advantages of treating individual spectral lines as independent RV estimators. Although primarily motivated by the treatment of outlying spectral information, this methodology applies directly to stellar activity and SPI. More recently, \citet{Martinez26} explored a similar strategy, reinforcing the idea that line-resolved measurements can provide a more physically informative view of stellar and planetary signals.

By avoiding the compression of the full spectrum into a single average profile, these methods retain information that may otherwise be lost in global RV estimators. This is particularly relevant for SPI studies, in which the signal may affect only specific lines or classes of lines rather than the spectrum as a whole.

\subsection*{Radial Velocity Activity Indicators}

The detection of exoplanets via RV is an indirect method, and from the earliest discoveries there has been concern about false-positive signals. This led to the development of activity indicators derived from the same spectra used to measure RVs. The most widely used of these is the bisector inverse span (BIS), popularized in the context of exoplanet detection by \cite{Queloz2001}. In this approach, the bisector of the CCF is measured in a manner analogous to that used for individual spectral lines in stellar physics \cite[e.g.][]{Gray05}.

Another important indicator is the full width at half maximum (FWHM) of the lines, often measured on the CCF. When derived from spectra obtained through a spectrograph with stabilized instrumental profile, variations in the FWHM primarily trace changes in the widths of spectral lines, and thus stellar activity. This has been used effectively to identify and model rotational modulation, for example in the ESPRESSO campaign of \cite{Faria22}. The line contrast can also serve as an activity proxy, alongside a broader set of indicators \citep[see, e.g.][]{Figueira13,Simola19,Barnes24}. However, these indicators often exhibit different sensitivities and correlations with RVs in active stars, as shown, for instance, by \cite{Lafarga20}, making their interpretation non-trivial.

In addition to line-shape diagnostics, wavelength-dependent indicators have been developed to probe activity-induced RV variations. Since the reflex motion induced by a planet is, to first order, achromatic, measurements of RVs as a function of wavelength have long been used to identify signals arising from stellar activity \citep[e.g.][]{Huelamo08}. In this context, several wavelength-dependent diagnostics have been developed to flag false positives, including the chromatic index (CRX) implemented in \texttt{SERVAL} \citep{Zechmeister18} and the \texttt{dTemp} indicator introduced by \citet{Artigau22}.

While these indicators can be straightforwardly derived from spectral lines or CCFs, they inherit the limitations associated with compressing the spectral information into a small number of scalar quantities, and may therefore miss line-dependent or chromatic signatures relevant to SPI.

\vspace{1.5cm}

Precise RV measurements and their associated diagnostics provide a rich framework for studying SPI. Methods based on global spectral compression, such as Fourier-domain approaches, CCFs, and template matching, have enabled the extraordinary precision required for exoplanet detection, but they remain fundamentally optimized for signals that affected by pure Doppler shifts. In contrast, SPI signatures are likely to manifest through subtle line-dependent, chromatic, or profile-shape variations that violate these assumptions. The more flexible and information-preserving approaches, including line-by-line analyses and data-driven decompositions of spectral variability draw us significantly closer to the objective of characterizing SPI. As a whole, the developments illustrate a broader transition in high-precision spectroscopy: from measuring a single scalar RV toward exploiting the full physical information stored in a stellar spectra.

\newpage

\section*{Precision Photometry from stellar surveys}


Precision time-series photometry provides a powerful avenue for studying SPI by enabling continuous monitoring of stellar brightness over timescales short enough to resolve stellar granulation signatures to those long enough to analyse stellar magnetic activity cycles. 
Unlike spectroscopic techniques, which probe localised diagnostics such as line cores or chromospheric emission, photometry measures the integrated response of the stellar photosphere. 
It is therefore sensitive to stellar activity across spatial scales, including starspots, faculae, flares, and transient occultations of active regions by orbiting planets \citep[see, for example][and references therein]{Huber25}.

From a methodological perspective, the primary challenge in photometric SPI studies is often not the detection of variability, but the robust separation of small, potentially planet-linked signals from other intrinsic stellar variability and instrumental systematics. 
Consequently, the effectiveness of photometry for SPI rests largely on advances in time-series analysis, correlated-noise modelling, and statistical inference \citep[e.g.][]{Cubillos17}.

Often, when we consider SPI, we are interested in photometric signals that occur with the same periodicity as that of an orbiting planet. 
To separate an SPI signal from these other factors, we then consider a useful conceptual decomposition of a photometric time series as
\begin{equation}
y(t) = m_{\star}(t) + m_{\mathrm{orb}}(t) + m_{\mathrm{tr}}(t) + s(t) + \epsilon(t),
\end{equation}
where $m_{\star}(t)$ represents intrinsic stellar variability, $m_{\mathrm{orb}}(t)$ any stellar component that occurs with the same period as the planetary orbit, $m_{\mathrm{tr}}(t)$ geometric apparent signals such as planet transits, $s(t)$ is any instrumental systematics, and $\epsilon(t)$ the residual noise. 
The methodological goal is to determine whether $m_{\mathrm{orb}}(t)$ is required by the data once realistic models of $m_{\star}(t)$ and instrumental effects are taken into account \citep[for more details, consider][]{Boisse11, Mcquillan12}

\subsection*{Separating Astrophysics from Instrumental Systematics}
Detrending is among the most consequential steps in photometric analyses of SPI. 
Many instrumental systematics, whether observations are collected from the ground or from space, occupy a similar frequency space as SPI signatures. 
The choice of how detrending is accomplished, or whether it is done at all, is a scientific choice that directly affects sensitivity, false-positive rates, and interpretability.
Effective detrending aims not to remove variability, but to separate components in a manner that preserves astrophysical signals while suppressing non-astrophysical ones, and effectively preserves information about the uncertainty inherent in the choices of detrending algorithm applied \citep[e.g.][]{Morello15, Hippke19, Taaki20}.

Two opposing failure modes dominate SPI analyses. 
\emph{Over-detrending} occurs when the detrending model is overly flexible and absorbs genuine astrophysical variability. 
This risk is highest when basis vectors or smoothing functions span timescales comparable to the stellar rotation or planetary orbital period, or when detrending is performed independently on short data segments, thereby destroying long-term phase coherence. 
Over-detrending biases results toward null detections and systematically underestimates signal amplitudes.

\emph{Under-detrending}, by contrast, leaves residual systematics that can masquerade as astrophysical signals. 
Low-frequency instrumental trends, cadence changes, and data gaps can imprint spurious periodicities or harmonics that align coincidentally with orbital phase. 
Under-detrending inflates false-positive rates and can lead to apparently coherent SPI signatures that disappear under alternative preprocessing choices.
Any result must demonstrate that it is not sensitive to the choice of detrending approach applied.

\subsubsection*{Families of Detrending Approaches}

Several methodological families are commonly employed to separate astrophysics from systematics.

\paragraph{Parametric and basis-vector detrending.}
Here, systematics are modeled as a linear combination of basis vectors,
\begin{equation}
s(t) = \sum_k c_k B_k(t),
\end{equation}
where $B_k(t)$ is inferred from low-rank trends derived from ensembles of stars or regressors tied to known instrumental states.
These may be found through an approach similar to principal component analysis, in which a set of orthogonal eigenvectors are calculated which parameterise the information shared across the entire ensemble.
This approach is efficient and interpretable.
However, it requires the basis vectors to not contain information with similar behavior to the astrophysical signals produced by the star. 
This was the primary approach of the ``pre-search data conditioning'' method of the \textit{Kepler} data analysis pipeline, which aimed to remove signals on both short and long timescales \citep{Smith12, Stumpe12, Stumpe14}

\paragraph{Filtering and smoothing.}
High-pass filters, running medians, or spline fits are often used to remove low-frequency trends.
One of the most common is a Savitsky-Golay filter \citep{Savitsky64}, which is a common ``flattening'' function in transiting planet analyses in which a low-degree polynomial is fit to a local window around each individual data point, and the calculated value of that polynomial at each point applied as the predicted model.
While simple, such filters impose an implicit transfer function on the data. 
This approach also assumes the noise properties of the data are unchanged throughout the window of observations, as each data point is given equal weight. 
The underlying data is also assumed to be characteristic of this smooth polynomial; any sharp features such as transits or stellar flares will bias this result.
In SPI applications, filtering must be chosen so that its characteristic cutoff lies well away from both the stellar rotation and planetary orbital periods; otherwise, genuine phase-coherent signals may be attenuated or distorted \citep[e.g.][]{Lanza08}.

\paragraph{Joint modeling of systematics and astrophysics.}
A more robust approach is to model systematics and astrophysical variability simultaneously, allowing uncertainty in their separation to propagate naturally into the final inference. \citet{Foreman-Mackey17} use 150 basis vectors rather than the standard 4-16 but simultaneously fit these with a transit model, reducing the risk of overfitting.

This joint formulation allows correlations between instrumental effects and stellar variability to be handled self-consistently, reducing biases that can arise when detrending and astrophysical modeling are performed sequentially. 
This approach is often implemented by combining low-dimensional parametric models or basis-vector representations for $s(t)$ with flexible stochastic models for $m_{\star}(t)$, such as Gaussian Processes, and explicit phase-locked components for $m_{\mathrm{orb}}(t)$. 
A key advantage of joint modeling is that uncertainty in the decomposition of variability is propagated into the posterior distributions of the SPI parameters, rather than being fixed at an earlier preprocessing stage. 
The principal drawbacks are increased computational cost, sensitivity to prior assumptions, and the need for careful model comparison to avoid overfitting, particularly when the systematics and astrophysical signals share similar characteristic timescales.
In practice, joint modeling is most useful when instrumental systematics and stellar variability operate on overlapping timescales, when the SPI signal of interest is expected to be low amplitude, or when the scientific conclusions depend sensitively on propagating detrending uncertainty into the final inference.

\subsection*{Modeling Stellar Variability as Correlated Noise}

Stellar photometric variability is neither white nor stationary: signals arising from starspots and faculae evolve in amplitude and phase as active regions grow, decay, and migrate across the stellar surface. 
For SPI studies, this variability is typically a nuisance signal that must be modeled and marginalized over to test for smaller, planet-linked effects. 
Gaussian Process (GP) regression has therefore become a widely used and practical tool for modeling stellar variability in photometric time series \citep[][and references therein]{Aigrain23}.

A GP can be understood as a flexible way of describing what stellar variability usually looks like without committing to a specific physical spot model. 
Instead of fitting individual spots, the GP builds models consistent with the expected covariance of the light curve: points close together in time are more similar than distant ones, and variability often repeats approximately on the stellar rotation period while evolving over longer timescales.

\newpage

A commonly used kernel for rotational modulation is the quasi-periodic form
\begin{equation}
k(t,t') = A^2 \exp\left[
-\frac{(t-t')^2}{2\lambda^2}
- \Gamma^2 \sin^2\left(\frac{\pi (t-t')}{P_{\mathrm{rot}}}\right)
\right],
\end{equation}
where $P_{\mathrm{rot}}$ represents the stellar rotation period, $\lambda$ the typical lifetime of active regions, and $A$ the overall variability amplitude. 
In practice, this kernel captures the intuitive picture of a light curve that is approximately periodic but slowly changing from one rotation to the next \citep{Angus18}.

In SPI analyses, the GP is rarely the quantity of interest itself. 
Rather, it serves as a flexible model for stellar variability that allows other components of the light curve to be tested more robustly. 
Potential SPI signatures are introduced as explicit orbital-phase terms, such as a sinusoid or low-order Fourier series at the planetary orbital period, and the question becomes whether the data prefer a model that includes both stellar variability and a phase-locked component over one that includes stellar variability alone.

A key strength of the GP approach is that it naturally accounts for correlated noise and propagates uncertainty in the stellar variability model into the inferred properties of any SPI signal. 
However, GPs are limited by the assumptions of the noise properties of the kernel itself, as well as the priors applied. In particular, if the GP contains periodic terms which are close in frequency to the planetary orbital period or its harmonics, a GP can absorb genuine phase-locked signals \citep[e.g.][]{Foreman-Mackey17}.

For this reason, GP-based analyses in SPI studies should be accompanied by explicit null tests and injection-recovery experiments to verify that orbit-locked signals of interest are not affected by the modeling procedure.

\subsection*{Flare Detection and Orbital-Phase Statistics}
Stellar flares appear in photometric time series as impulsive, asymmetric brightenings with a rapid rise and more gradual decay. 
In the context of SPI studies, flares are of particular interest because magnetic interactions between a star and a close-in planet have been proposed as a possible trigger or modulator of flare activity \citep{Ilin25}.
To test this, we then must be able to reliably identify flares in survey photometry and test whether their occurrence shows any preference for particular planetary orbital phases\footnote{See, however, the different flare signature triggered by magnetic SPI and proposed for GJ\,436 by \cite{Loyd2023}.}.

Flare detection typically begins with the removal of low-frequency variability associated with stellar rotation and instrumental trends. This step is intended to flatten the local baseline without suppressing impulsive events \citep[e.g.][]{Davenport16}.
Candidate flares are then identified as statistically significant upward excursions above the local noise level, using a simple flare template \citep{Notsu13}, or sometimes a machine learning approach \citep{Vida18, Feinstein20}.

Once flares are identified, each event time is converted into an orbital phase, in order to test  whether flares preferentially cluster at specific orbital phases. This typically involves randomizing flare times within the observed time windows to estimate how often comparable clustering would arise by chance \citep[e.g.][]{Tofflemire17, Ilin25}. The phase conversion is done relative to a reference time T$_0$ (for example the transit midpoint), together with either the orbital period $P_{\mathrm{orb}}$ or the synodic period $P_{\mathrm{syn}}$.

In most datasets, flare detectability is not uniform in time. 
Variations in noise level, data gaps, and detrending efficiency all affect the probability of detecting a flare at a given epoch, and must be accounted for in data analyses.
This is further compounded by intrinsic stellar variability, since stars with similar spot coverage can exhibit markedly different flare occurrence rates, likely reflecting differences in magnetic complexity rather than spot area alone \citep{Araujo2023}.

\subsection*{Mapping Stellar Surfaces}

For transiting systems, precision photometry provides a unique opportunity to probe stellar surface structure through the analysis of starspot and facular crossings. 
Spots and faculae produce brightness variations on a rotational timescale \citep[e.g.][]{Berdyugina05, Oshagh13}.
When a planet occults an active region on the stellar surface, the resulting transit light curve deviates locally from the smooth profile expected for a uniform photosphere \citep{Silva03}.
These anomalies appear as short-duration brightenings or dimmings within transits and offer spatially resolved information about stellar activity that is otherwise inaccessible in disk-integrated photometry \citep{Morris17}.

This analysis typically begins with fitting a standard transit model to each individual transit event, with residuals inspected for statistically significant localized features. The timing, duration, and amplitude of these events is sufficient to infer their approximate position,  size, and intensity along the transit chord \citep{Silva-Valio2011, Zaleski2022, Araujo2023}

By converting the timing of each anomaly into a position on the stellar surface, spot-crossing events can be mapped to stellar longitudes \citep{Lanza09, Valio2017}.
Repeated transits allow these active regions to be tracked over time, providing constraints on spot lifetimes, differential rotation \citep{Silva-Valio2011, Valio2017, Araujo2021, Zaleski2022, Valio2024}, and the persistence of preferred active longitudes \citep{Nutzman11, Sanchis-Ojeda11}.
When spot-crossing events recur at similar stellar longitudes over many transits, they indicate long-lived surface structures. 
Moreover, transit spot mapping enables the estimation of starspot magnetic fields through calibrated relations between spot flux deficit and magnetic field, yielding typical values of $\sim 2.7–4.6$ kG across FGK and M stars \citep{Menezes2024}. In parallel, it provides robust measurements of spot filling factors, with surface coverage ranging from a few percent up to $\sim30$\%, revealing that the decline of stellar magnetic activity with age is primarily driven by a reduction in spot coverage rather than in field strength \citep{Araujo2025}.

For SPIs, a primary question is often whether such active regions show a systematic relationship with the planetary orbit. This may manifest as spot crossings occurring preferentially at specific orbital phases or longitudes tied to the planet's position \citep{Shkolnik08, Pagano09}.

Several limitations must be borne in mind when interpreting spot-crossing signatures. 
The stellar surface properties remain ambiguous when a transit is not occurring because of a loss of information as a stellar surface is projected into a 1-dimensional time series \citep{Luger21, Deagan26}.
Small or low-contrast active regions may escape detection, leading to incomplete surface maps. 
Degeneracies between spot size, contrast, and latitude can complicate physical interpretation, particularly for grazing transits. 
In addition, apparent recurrence of anomalies can arise spuriously if the stellar rotation period is close to an integer multiple of the orbital period.

\subsection*{Statistical Validation and Sensitivity Limits}

Most photometric SPI signals are subtle and require statistical validation; many studies are based on population-level inference rather than analysis of a single star.
Statistical significance must be evaluated relative to realistic null models. 
These are rarely white noise, but rather include evolving stellar variability, correlated noise, and complicated observing window functions. 
As a result, these models are often estimated empirically using techniques such as orbital-phase scrambling, time-shifting of the ephemeris, or bootstrap resampling that preserves the temporal structure of the data while destroying long-term phase coherence.

Injection-recovery experiments play a dual role in this context. 
Beyond validating detrending, they define the sensitivity of the analysis to SPI-like signals. 
By injecting synthetic orbital-phase modulations, clustered flare populations, or spot-crossing anomalies across a grid of amplitudes and phases, one can construct detection-efficiency curves that quantify the probability of recovery given the data quality and analysis choices \citep{Christiansen16}. 
In the absence of a detected signal, these curves provide robust upper limits on the amplitude or rate of any planet-linked photometric variability \citep{Miller12}.

Population-level analyses offer an additional layer of statistical power \citep{Mazeh15}. 
By comparing ensembles of stars with and without close-in planets, or by stacking phase-folded signals across multiple systems, marginal effects that are undetectable in individual light curves can be identified statistically.
Such approaches require careful matching of control samples and explicit treatment of selection effects, but they can provide strong constraints on the prevalence of SPI \citep[e.g.][]{Ilin24}.

\newpage

\section*{Chromospheric activity indexes}


Chromospheric spectral lines are arguably the most widely used diagnostics of stellar magnetic activity \citep[e.g.][]{Hall08} and have been prime targets in the search for SPI signatures. These lines are formed in the upper layers of the stellar atmosphere, where non-LTE conditions and magnetic energy dissipation become non-negligible \citep[e.g.][]{Linsky17}. The emission in the line cores is ultimately linked to the excitation and ionization of atoms in plasma whose thermodynamic state is strongly modulated by the local magnetic field. For this reason, chromospheric lines provide a more direct probe of magnetically driven variability than RVs, line-shape diagnostics, or photospheric observables. \\

We review the spectral lines and methods traditionally used to quantify chromospheric activity, and then consider how these diagnostics are adapted or reinterpreted in the context of SPI.

\subsection*{Ca \scriptsize{II}\normalsize\- H \& K lines}

The most well-known chromospheric activity indicator is the Ca {\scriptsize{II}} H \& K doublet. It was used to study the long-term activity evolution of solar-neighbourhood stars in the foundational work of \cite{wilson1978}. The two lines are located at 393.3664\,nm and 396.8470\,nm and can be detected from a spectral resolution of R$\sim$10\,000 due to their intrinsic width. Since RV planet-hunting spectrographs typically have R$\geq$60\,000 and cover (traditionally) the visible range, one is capable of measuring these lines from RV campaign spectra, allowing an activity characterization contemporaneous with RV measurements \citep[e.g.][]{Boisse11}. 

The activity index log(R'$_{\mathrm{HK}}$) defined by \cite{Noyes84} remains in widespread use today, which attests to its effectiveness. Several teams have adapted the index to use in modern instrumentation \citep[e.g.][]{Lovis11} or to extend to M-dwarf stars \citep[e.g.][]{SuarezMascareno15}, but its underlying principles remain essentially unchanged. The flux in the line cores is measured and normalized by the nearby continuum. This ratio is then calibrated as a function of spectral type (often via $B-V$), in order to remove the photospheric contribution and yield a value that can be compared across stars of different spectral class.

\begin{figure}[h]
 \centering
   \includegraphics[scale=0.7]{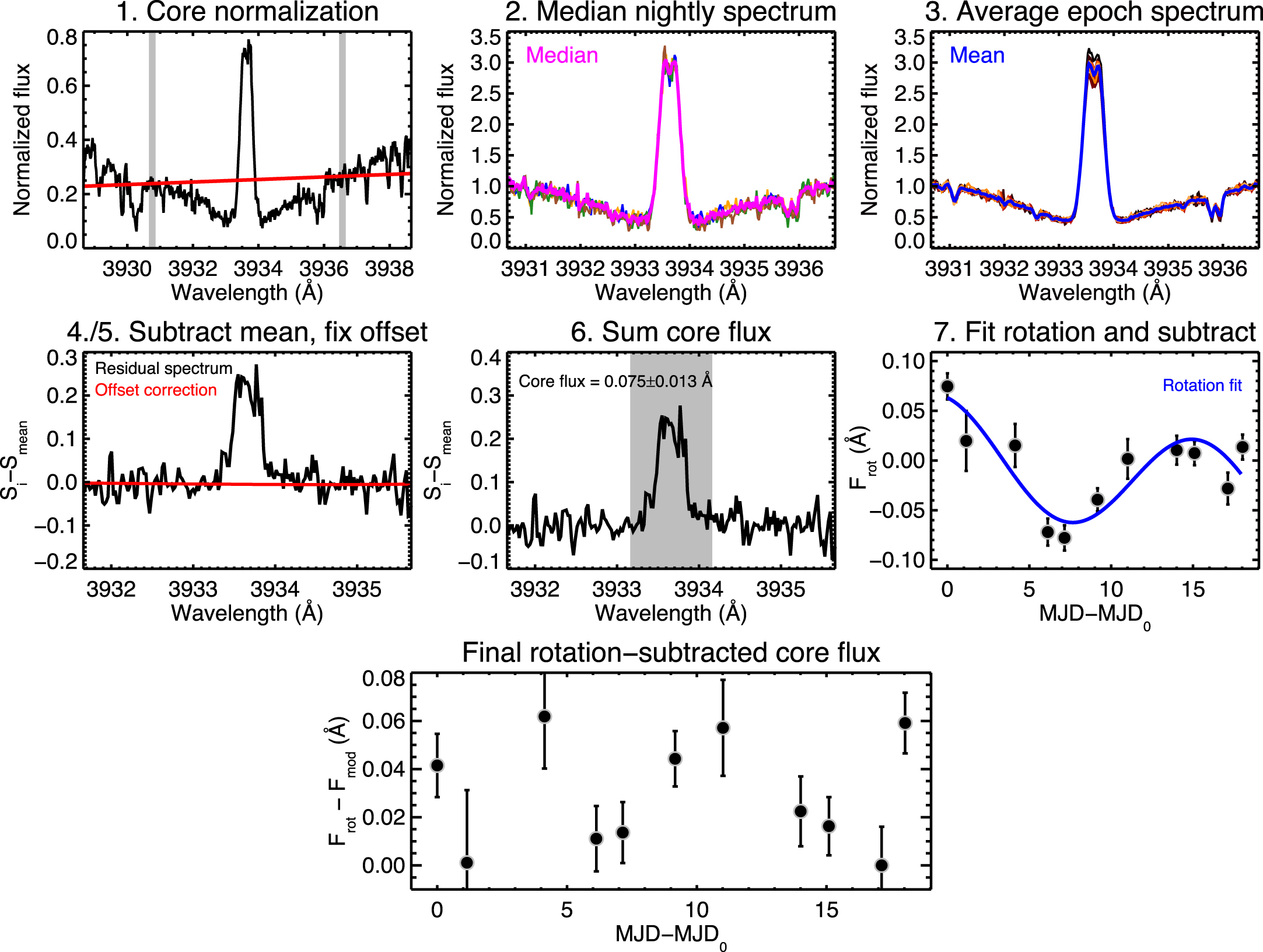}
\caption{The derivation of the Ca \scriptsize{II}\normalsize\- K flux, from \cite{Cauley2018}. See text for details.}
\label{Fig:Cauley}
\end{figure}

Reported SPI detections, so far, have been based on detections in the Ca \scriptsize{II}\normalsize\- lines. In the context of these studies, the analysis of the lines is done differently from the traditional log(R'$_{\mathrm{HK}}$) described above. In most SPI studies, the H and the K lines are treated separately; no index is calculated, but a variability measurement is done directly in the line core. For that, a bandpass of 7\AA\ is centred around each line core. In order to normalise the spectrum, the ends of the bandpass are set to 1 \cite[see, e.g.][]{Shkolnik03,Shkolnik05,Shkolnik08}. An average profile is then calculated from all observations, and the residual profile per observation is calculated by subtracting the average profile from the line core (within the bandpass). In principle, the residual profiles should only show variability in the core of the Ca line, and the edge of these profiles should be zero (within their error bars). In order to make sure that only the variability in the core is considered, some authors have corrected the residual profiles from possible continuum variability by subtracting a low order polynomial from the residuals \citep[an offset correction as defined in, e.g.][]{Cauley2018}. Once the final residuals profiles are calculated (whether the last correction step is added or not), the flux in the core of the line is measured, and its variability is investigated for modulation with periods associated to SPI. Figure \ref{Fig:Cauley} summarises the steps followed for the calculation of the index.
We note here that some studies have combined the information from both the Ca \scriptsize{II}\normalsize\- H \& K lines \citep[see, e.g.][]{fares2009,fares2010,fares2012,klein2022}, while others used the traditional index definitions.

A very recent study of the warm Neptune GJ\,436\,b reports evidence for magnetically-driven SPI traced through the Ca \scriptsize{II}\normalsize\ H \& K lines and the Ca infrared triplet \citep{2026arXiv260406969R}. The authors identify a periodic modulation of the host star’s activity on the timescale of the planet’s synodic and anti-synodic periods on top of the stellar rotation one. The signal is independently recovered with both HARPS and CARMENES, and shows an on–off behaviour, being detectable when the star is in an intermediate activity state of its long-term activity cycle. This result provides a rare indirect constraint on the magnetic field of a Neptune-mass exoplanet and extends SPI studies beyond the hot-Jupiter population.

\subsection*{H${\alpha}$}

The H$\alpha$ line (located at 656.28 nm) is a fundamental diagnostic of chromospheric magnetic activity, particularly in late-type stars, where its emission strength and profile changes reflect non-thermal heating processes associated with features such as plages, filaments, and flares \citep{Linsky17,Fuhrmeister19,kuridze2015}. 
Although both H$\alpha$ and the Ca \scriptsize{II}\normalsize\- H\&K resonance lines are primary diagnostics for stellar magnetic activity, their inter-relationship is complex and frequently reveals significant discrepancies. These differences arise because the two tracers probe distinct atmospheric heights: Ca \scriptsize{II}\normalsize\- H\&K form in the middle-to-lower chromosphere, while H$\alpha$ originates in the upper chromosphere \citep{GomesdaSilva11,Linsky17}. Consequently, they respond differently to magnetic structures; while \textit{plages} generally produce a positive correlation in both lines, \textit{filaments} can introduce negative or anti-correlated signals in H$\alpha$ \citep{Meunier22}. 

Beyond these discrepancies, one advantage of the H$\alpha$ line is its location in the red part of the optical spectrum, where it is generally easier to observe at high S/N than Ca \scriptsize{II}\normalsize\- H \& K, particularly in the cool M dwarf stars that became the focus of many RV surveys. This made the line an attractive activity tracer in RV campaigns and, by extension, in SPI studies. As with Ca \scriptsize{II}\normalsize\- H \& K, the usual approach is to measure the flux in the line core relative to nearby continuum bands, thereby defining an H${\alpha}$ activity index. The bandwidth used to calculate the H$\alpha$ index is critical for its agreement with Calcium. In solar-type stars, an optimized 0.6 Å bandwidth focuses on the core emission and maximizes the positive correlation with Ca \scriptsize{II}\normalsize , whereas the standard 1.6 Å filter incorporates variability from the line wings that can mask rotational signals or generate artificial ``anti-cycles" \citep{GomesdaSilva22}. For M dwarfs, the discrepancies are further accentuated because the Ca \scriptsize{II}\normalsize\- flux is often extremely low and noisy in the blue spectral region \citep{Fuhrmeister19,Meunier2024}.

The interpretation of H${\alpha}$ is often more nuanced than log(R'$_{\mathrm{HK}}$). Its response to magnetic activity depends appreciably on several parameters beyond spectral type, to the extreme that the line may appear either in absorption or emission depending on the physical conditions in the chromosphere. In low-activity stars, H$_{\alpha}$ may show weak or even no detectable correlations with Ca \scriptsize{II}\normalsize\-  H \& K, whereas in more active stars it can become a sensitive tracer of plages, flares, and transient chromospheric variability \citep[e.g.][]{Cincunegui07,GomesdaSilva11,IbanezBustos23}. Specifically in M dwarfs, H$\alpha$ activity—quantified by its fractional luminosity ($L_{H\alpha}/L_{bol}$)—exhibits a uniquely non-linear evolutionary path. As rotation rates and activity levels increase, the line first deepens in absorption before ``filling in" and eventually transitioning into full emission \citep{Newton17}. During high-energy flare events, this transition from absorption to emission can occur rapidly \citep{namekata2022}. Such sensitivity to both quiescent and transient chromospheric states necessitates extreme caution when interpreting orbit-phased or intermittent modulation as evidence of SPI.

\subsection*{He {\scriptsize{I}}}

The He\,{\scriptsize{I}} line most commonly used as an activity diagnostic in optical spectra is the He {\scriptsize{I}} D$_3$ triplet, located near 587.6\,nm. In cool stars, this line forms in the upper chromosphere and lower transition region, and its appearance is closely linked to non-radiative heating and high-energy irradiation \citep{Andretta1995}. Unlike traditional resonance lines such as Ca\,{\scriptsize{II}} H \&K or the H$\alpha$ Balmer line, the formation of the He\,{\scriptsize{I}} D$_3$ triplet in late-type stars is primarily driven by a  photoionization-recombination (PR) mechanism \citep{Andretta1995,Sanz-Forcada08}. 

Consequently, He\,{\scriptsize{I}} serves as a more selective tracer of energetic atmospheric processes than traditional indicators. While Ca\,{\scriptsize{II}} and H$\alpha$ reflect the bulk properties of the mid-to-lower chromosphere, and upper chromosphere, respectively, He\,{\scriptsize{I}} acts as a direct proxy for the high-energy radiation field and the presence of highly localized heating events \citep{Sanz-Forcada08,Guilluy20}. In practice, the line is typically quantified through indices based on its core flux or equivalent width (EW) measured relative to the local continuum. However, the observational implementation of He\,{\scriptsize{I}} as a tracer remains demanding. In inactive stars, the line is inherently weak and frequently contaminated by telluric absorption features or blended with nearby metallic lines, necessitating high-resolution spectra and meticulous data reduction \citep{Guilluy20,Fuhrmeister19} for a proper analysis. Furthermore, its response is non-linear; the line can appear in absorption, ``filling- in", or transitioning into emission depending on the specific thermal structure and radiation field of the upper atmosphere \citep{Andretta1995}. Despite these complexities, its unique sensitivity to the most energetic layers of the stellar atmosphere makes it exceptionally attractive for studying Star-Planet Interaction (SPI). In systems with close-in exoplanets, magnetic or particle-driven interactions—such as reconnection events or the formation of bow shocks—preferentially impact the upper chromosphere, potentially creating sub-orbital signatures in He\,{\scriptsize{I}} that remain distinct from the background stellar activity \citep{Guilluy20}.

\subsection*{Na {\scriptsize{I}}}

The Na {\scriptsize{I}} D doublet, located at 588.9950\,nm and 589.5924\,nm, is also commonly used as an activity diagnostic in cool stars. Although these lines are primarily photospheric in origin, their deep, narrow cores are formed in the lower-to-middle chromosphere, making them sensitive to non-thermal heating and magnetic activity \citep{Andretta1995,Montes1997}. In active stars, this sensitivity manifests as a central emission within the absorption cores, a process that can be quantified through activity indices defined by measuring the core flux relative to nearby continuum passbands, analogous to the methods used for Ca {\scriptsize{II}} H\&K and H$\alpha$ \citep{Diaz07}.

A primary observational advantage of the Na {\scriptsize{I}} D lines is their location in the yellow-red portion of the optical spectrum. For cool M dwarfs, which exhibit drastically reduced flux in the blue-UV region, the Na I doublet provides a high signal-to-noise (S/N) alternative to the traditional Ca {\scriptsize{II}} H\&K \citep{Diaz07,Fuhrmeister19}. This makes the doublet a staple in large-scale radial velocity (RV) surveys and long-term activity monitoring programs where blue-channel sensitivity may be limited. However, the interpretation of Na {\scriptsize{I}} activity is rarely straightforward \citep{Meunier2024}. The lines retain a substantial photospheric contribution, which must be carefully accounted for to isolate the chromospheric signal. Furthermore, the doublet is frequently affected by telluric water vapour absorption and interstellar medium (ISM) contamination; the latter is particularly problematic as ISM absorption features can overlap with the stellar line cores depending on the target's radial velocity \citep[e.g.,][]{Welsh2010,Diaz07}. 

In Star-Planet Interaction (SPI) studies, Na {\scriptsize{I}} provides a critical complementary diagnostic. Because it probes a transition region between the upper photosphere and the lower chromosphere, it offers a different perspective on the atmospheric response to planetary perturbations compared to the higher-forming H$\alpha$ or He {\scriptsize{I}} lines. This mixed sensitivity allows for a more comprehensive ``tomography" of the stellar atmosphere during suspected SPI events \citep[e.g.,][]{Cauley2018}. Nevertheless, the complexity of its formation requires caution, as observed modulations in the Na {\scriptsize{I}} index may reflect a combination of photospheric spots and chromospheric plages rather than a purely planet-induced signal \citep{klein2022,IbanezBustos23}.

\vspace{1cm}

Taken together, these diagnostics probe different layers and physical regimes of the stellar atmosphere, from the lower chromosphere to the upper chromosphere and transition region. Their simultaneous use is therefore particularly valuable in SPI studies, where the expected signal may be weak, intermittent, and manifest differently depending on the atmospheric depth and physical mechanism involved. Unfortunately, the characteristics of the star and properties of the spectra may make a multi-index analysis impractical.

\vspace{2.0cm}

\section*{Radio observational diagnostics of magnetic SPI}

Radio searches for magnetic star--planet interaction (SPI) have evolved from the initial, simple flux-density searches for exoplanetary cyclotron emission in the early 90s to multi-dimensional searches based on polarization, time--frequency behaviour, planetary orbital phase, and stellar activity context.
The physical basis remains the Solar-System analogy established by \citet{Zarka2007}: a plasma flow interacting with a magnetized or conducting obstacle can drive electron acceleration and coherent electron-cyclotron maser (ECM) emission. In direct planetary emission, the relevant magnetic field is the planetary field and the maximum frequency is
\begin{equation}
  \nu_{\rm c} \simeq 2.8\,B\;{\rm MHz},
\end{equation}
where $B$ is in Gauss. This immediately imposes a strong observational selection effect: terrestrial planets with Gauss-level fields emit below or close to the ionospheric cut-off, whereas hot Jupiters or strongly magnetized planets may reach the decametre--metre window. In sub-Alfv\'enic SPI, however, the emission can be produced in the stellar magnetosphere, so the relevant field is the stellar coronal field. For active M dwarfs, with surface fields of hundreds of gauss to kilogauss, the predicted ECM frequencies move into the LOFAR/uGMRT/VLA/ATCA range, from tens of MHz to several GHz \citep{Zarka2007,Saur2013,Turnpenney2018,Vedantham2020,PerezTorres2021,Callingham2024}.

The main observational advance has therefore been methodological rather than simply instrumental. A credible radio SPI experiment now requires simultaneous or sequential tests of: (i) coherent-emission diagnostics, especially high circular polarization and high brightness temperature; (ii) time-domain behaviour, including bursts, persistence, duty cycle, and dynamic spectra; (iii) frequency placement relative to plausible cyclotron frequencies; (iv) phase coherence with the orbital or synodic period; and (v) rejection of ordinary stellar activity through chromospheric, coronal, rotation, and flare information. None of these criteria is individually decisive, but together they define the current evidentiary standard for candidate magnetic SPI.

\subsection*{From targeted low-frequency searches to Stokes-$V$ survey selection}

The first generation of exoplanet radio searches was motivated by the radiometric Bode-law and radio-magnetic scaling-law extrapolations from Solar-System planets \citep{Zarka2007,Zarka2018}. Observations targeted systems expected to maximize incident stellar-wind power, planetary magnetic moment, or proximity, often at low frequencies where Jovian analogues should be brightest. The observational limitations were severe: ionospheric access, radio-frequency interference, uncertain beaming, poor knowledge of planetary magnetic fields, and the possibility that the stellar wind or planetary plasma environment suppresses ECM escape \citep{Turner2021,Callingham2024}. Consequently, many searches have produced upper limits rather than detections, and non-detections were difficult to interpret uniquely.

The advent of instruments working at very low frequencies and with enormous sensitivities, such as LOFAR, enabled sensitive, blind wide-field imaging in both total intensity and circular polarization rather than pointing only at pre-selected exoplanet systems. \citet{Vedantham2020} identified GJ~1151 by cross-matching LoTSS sources with nearby Gaia stars. The source was detected at 120--167 MHz with a high circularly polarized fraction, persistent emission over an $\sim 8$ h observation, and a quiescent chromospheric activity state. Its brightness temperature and polarization argued against incoherent gyrosynchrotron emission, while plasma emission was disfavoured by the combination of polarization and coronal-density constraints. The remaining interpretation was auroral ECM powered by a large-scale current system, plausibly a sub-Alfv\'enic interaction with a short-period planet \citep{Vedantham2020}. However, since no close-in planet has been detected in the system, nor has LOFAR confirmed the emission in subsequent observations, the SPI interpretation remains simply a possibility.

This approach was generalized by \citet{Callingham2021}, who used LoTSS Stokes-$V$ maps and Gaia cross-matching to build a blind sample of 19 low-frequency M-dwarf detections. Most sources displayed high circular polarization, brightness temperatures above $10^{12}$ K, broad effective bandwidths, and hour-scale persistence. The important methodological result was the population comparison: radio luminosity did not correlate cleanly with standard chromospheric or coronal activity indicators, and slowly rotating, X-ray faint stars became the cleanest SPI candidates. Thus, survey selection in Stokes $V$ provides a way to find systems where the radio emission is not simply tracing ordinary flare activity, although it does not by itself prove the presence of a planet.

\subsection*{Targeted monitoring of known planet hosts}

A complementary method is phase-resolved monitoring of known close-in planet hosts. The most developed example is the ATCA campaign on Proxima Centauri by \citet{PerezTorres2021}. The experiment used 18 consecutive daily observations from 1.1 to 3.1 GHz, covering about 1.6 orbital cycles of Proxima~b. The observing frequency was chosen because the stellar magnetic field of Proxima, $B_\star \simeq 600$ G, implies an ECM frequency near 1.7 GHz if the emission arises in the stellar magnetosphere. The analysis combined full-Stokes imaging, frequency-resolved flux measurements, short-timescale searches in the visibility domain, and folding of the radio light curve on the known planetary orbit. The key diagnostics were the strong low-frequency component around 1.6 GHz, high circular polarization, reversals of the Stokes-$V$ sign, short flares, and a multi-day burst with maxima close to the quadratures of Proxima~b. The interpretation remains a candidate rather than a confirmed detection, but it established the template for GHz-frequency SPI monitoring: select the band from the stellar magnetic field, monitor long enough to test orbital phase, and use polarization plus spectral cut-offs to discriminate ECM from plasma emission or gyrosynchrotron radiation.

The same logic has been applied to other nearby M-dwarf planet hosts. \citet{PinedaVilladsen2023} reported 2--4 GHz coherent bursts from YZ~Ceti, with repeated bursts occurring at similar orbital phases of YZ~Ceti~b, consistent with enhanced burst probability near the predicted SPI phase, while retaining caution about intrinsic stellar bursts. \citet{Trigilio2023} used uGMRT Band~4 observations of YZ~Ceti and argued, from repeated detections, circular polarization, and orbital-phase statistics, for auroral emission associated with SPI. These works show why phase-resolved monitoring is now central: a single polarized burst demonstrates coherent emission, but periodic recurrence at the planet's orbital or synodic phase is what turns a stellar-radio detection into an SPI candidate.

Non-detections are equally important when they are obtained with complete phase coverage and appropriate Stokes-$V$ sensitivity. \citet{Narang2024} re-observed GJ~1151 with the uGMRT at 150, 218, and 400 MHz and did not recover the LOFAR-like emission, emphasizing that SPI candidates may be intermittent, beamed away from the observer, absorbed, or active only under favourable stellar-wind conditions. \citet{PenaMonino2025a} carried out nine uGMRT epochs of GJ~486 from 550 to 750 MHz, covering almost the full orbital phase of GJ~486b. The observations produced no Stokes-$I$ or Stokes-$V$ detection and no burst in dynamic spectra. Because the band was selected from the expected stellar ECM frequency, the non-detection could be translated into constraints on the stellar mass-loss rate, radio conversion efficiency, and beaming geometry. This illustrates a newer use of radio SPI observations: even non-detections can constrain the allowed parameter space if the campaign has adequate orbital coverage, polarization products, dynamic spectra, and a forward model including the stellar wind, magnetopause size, and free--free absorption.

\subsection*{Dynamic spectra, polarization, and mechanism discrimination}

The core data product for SPI searches is no longer only a continuum image but a set of Stokes-$I$ and Stokes-$V$ images, light curves, and dynamic spectra. Imaging is needed to verify positional association and remove unrelated background sources. Stokes $V$ is crucial because coherent ECM can approach very high circular polarization, whereas most incoherent stellar emission is weakly polarized. Dynamic spectra are needed because ECM may appear as narrowband or structured bursts that are diluted in time-averaged imaging. They also allow searches over several time and frequency binnings, which is essential when the intrinsic burst duration and bandwidth are unknown \citep{PenaMonino2025a}.

The main ambiguity is between ECM and coherent plasma emission. Plasma emission can also be bright and polarized, particularly in active coronae, but it is tied to the plasma frequency and its harmonics, depends strongly on coronal density, and often has different polarization and time--frequency behaviour. ECM is tied to the cyclotron frequency, requires low plasma-to-cyclotron frequency ratios along the escape path, and is expected to be beamed. Practical discrimination therefore uses a combination of observables: very high brightness temperature, high circular polarization, spectral cut-offs consistent with $\nu_{\rm c}$, stable polarization handedness or hemisphere-dependent reversals, hour-scale auroral persistence, and recurrence at orbital phase \citep{Vedantham2020,PerezTorres2021,Callingham2021,Callingham2024}. Ancillary optical, H$\alpha$, X-ray, rotation, and magnetic-field measurements are required to decide whether the source is a normal active star, a rotation-powered auroral emitter, or a plausible planet-induced system.

\subsection*{Current best practice and remaining limitations}

The current best practice for radio SPI searches is therefore a matched observing and modelling workflow. First, targets are selected among nearby M dwarfs or known close-in planet hosts for which the expected cyclotron frequency is accessible and the planet may orbit inside the Alfv\'en surface. Second, observations are designed to cover the relevant orbital and synodic phases, ideally over multiple cycles, with full-Stokes products and sufficient time--frequency resolution to form dynamic spectra. Third, candidate detections are evaluated against a hierarchy of tests: positional coincidence, high Stokes-$V$ fraction, brightness temperature, spectral behaviour, recurrence, orbital/synodical-phase clustering, and absence of ordinary stellar-activity counterparts. Finally, detections and non-detections are interpreted with forward models of the sub-Alfv\'enic interaction, beaming, stellar magnetic geometry, stellar wind density, and propagation losses \citep{Saur2013,Turnpenney2018,Callingham2024,PenaMonino2025b}.

The field has not yet reached the level of an unambiguous, repeatable radio detection of magnetic SPI. The leading candidates show several of the expected signatures, but the evidence is limited by beaming, intrinsic intermittency, stellar flare contamination, incomplete knowledge of the stellar wind, and sparse repetition of orbital phase coverage. The methodological progress is nevertheless substantial. Radio SPI searches have become quantitative tests of magnetospheric electrodynamics rather than simple source searches, and the combination of LOFAR, uGMRT, VLA/ATCA monitoring, optical radial velocities, magnetic mapping, and future LOFAR~2.0/SKA-Low sensitivity should make it possible to distinguish persistent stellar aurorae, stochastic coherent flares, direct planetary aurorae, and genuine planet-induced radio emission.

\newpage

\section*{Transmission spectroscopy}

The interaction between a planet and its host star leaves various imprints on the planet's atmosphere. These interactions are driven by processes that include, mainly, the deposition of
stellar radiation on the planet, the interception of the planetary wind by its stellar counterpart, and the coupling of the planetary outflow with the star-planet magnetic environment. 
Most efforts to characterize exoplanet atmospheres take advantage of such interactions which, under appropriate conditions, make the atmospheric signatures more readily detectable. 
Furthermore, there is an increasing interest in using a few specific atmospheric tracers to constrain the stellar properties, a goal that requires a sound understanding of the 
processes behind star-planet interactions.\\


Among the techniques used for the characterization of exoplanet atmospheres, transmission spectroscopy has been the most successful to date. The atmosphere is revealed through the difference in the joint star-planet spectra measured 
while the planet is transiting its host star relative to  the spectrum while the planet is off transit. 
Because the transit probability is inversely proportional to orbital distance, transmission spectroscopy is best suited for the study of close-in exoplanets. 
This constraint prevents its use in the characterization of the vast majority of exoplanets, including Proxima Centauri b -- the nearest to us, and which orbits the habitable zone (HZ) of its host star \citep{Anglada-Escude16,jenkinsetal2019}. 
On the other hand, the proximity of most star-planet systems studied using transmission spectroscopy often leads to enhanced star-planet interactions and stronger atmospheric signatures. 
The signatures of the chemical species searched for with transmission spectroscopy
will be strong if the species are abundant and their cross sections large. 
For example, the cross sections of dipole-allowed atomic lines can be as large as 10$^{-13}$ cm$^2$ at the line's core, enabling the detection of major species out to far distances from the planet, and of minor species to altitudes similar to the planet's optical radius. By targeting multiple lines of the same or different species, transmission spectroscopy renders possible the investigation of star-planet interactions over a range of conditions and altitudes in the planet's atmosphere. In short, transmission spectroscopy is a unique tool to explore the composition, temperature and dynamics of an atmosphere and their connection with the stellar environment.
\\ 

To illustrate these possibilities, we focus on the {\hi} {\lalpha} line at 1,216 {\AA}, the {\hi} Balmer, the {\cii} resonance line at 1,335 {\AA} and the {\hei} triplet line at 1.08 $\mu$m. They trace the low-density atmospheric layers in the upper atmosphere where the interaction with the star is stronger. 
For the gas to reach such altitudes, and possibly escape the planet's gravitational pull, the planet must receive large amounts of stellar radiation. Stellar XUV photons (of wavelengths less than the Lyman continuum threshold at 912 {\AA}) are crucial to photoionize the {\hi} atoms, often prevalent at high altitudes, and heat the atmosphere as the ejected primary electrons transfer their kinetic energy to the gas \citep{cecchi-pestellinietal2006,garciamunoz2023}. 
Unfortunately, the XUV spectra of most stars are poorly known because it is masked by the interstellar medium. 
Much work is being done to fill in this critical gap, which often requires extrapolations based on reference stars or the reconstruction of the stellar corona from measurements of the stellar flux at X-ray and FUV wavelengths \citep{youngbloodetal2016,wilsonetal2025}.
The detection of these lines requires moderate-to-high spectral resolution, typically 
$\lambda$/$\Delta \lambda${$>$}1,000
and a bright stellar background. Both requirements are more easily met at visible and infrared wavelengths, as many medium and large-size ground-based telescopes are equipped with high-resolution spectrographs, like CRIRES+, NIRPS, CARMENES, GIANO and SPIRou.
For lines at wavelengths shorter than the ozone cutoff of {$\sim$}3,100 {\AA}, observing from space is a must. Indeed, UV transmission spectroscopy of exoplanets has mostly relied on the Hubble Space Telescope. 
Ensuring the availability of space telescopes capable of UV spectroscopy of exoplanets
to build upon HST's legacy should remain a priority. The proposed Habitable Worlds Space Observatory and other international ventures \citep{gomez-de-castroetal2022,cubillosetal2025,dossantosetal2025} will supersede HST in the mid-term.\\

In absorption, the {\hi} {\lalpha} line  arises in the dipole-allowed transition H(1$s$)+$h\nu$(1,216 {\AA}){$\rightarrow$}H(2$p$). The abundance of {\hi} atoms in the atmospheres of the close-in gas giants targeted in early transmission spectroscopy surveys led to the rapid detection of the line \citep{vidal-madjaretal2003}.
On the well-known hot Jupiter HD 209458 b, the line remains opaque over at least $\sim${$\pm$}100 km/s from line core, that corresponds to a projected area equivalent to a few times the planet's optical radius \citep{vidal-madjaretal2003,linskyetal2010}. Models indicate that hydrogen is lifted to such high altitudes by strong XUV irradiation from its Sun-like host star \citep{yelle2006,garciamunoz2007}. 
At the also well-known hot-Jupiter HD 189733 b,  absorption in the {\hi} {\lalpha} line has been reported to fluctuate over time \citep{lecavelierdesetangsetal2010,lecavelierdesetangs2012}, becoming strong at some epochs while going undetected at others. This temporal variability in the planet's atmosphere is likely an outcome of variability in the radiative output or the wind properties of the host star \citep{cherenkovetal2017,hazraetal2022}. 
This connection offers a means of probing the stellar properties, including the magnetic field \citep{khodachenkoetal2021}, from their effects on the planet atmospheres. 
For these mechanisms to be truly effective and make a detectable impact, however, they must last at least a time of a few hours comparable to that of the transit duration \citep{cherenkovetal2017,hazraetal2022,gilletetal2025}.
\\ 

\begin{figure}
   \centering
   \includegraphics[width=16.cm]{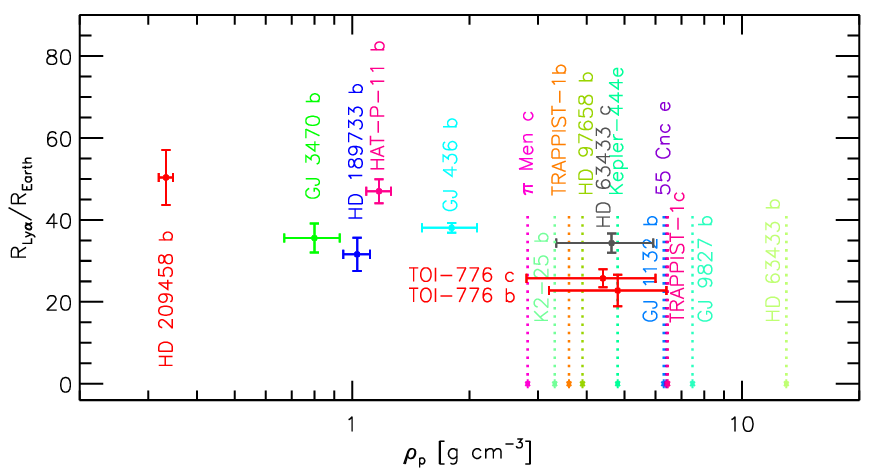}   
      \caption{\label{lymanalpha_fig} 
Summary of detection attempts of {\hi} Lyman-$\alpha$ absorption on exoplanets. Figure based on \citet{garciamunozetal2020}, and updated with recent observations. The measured transit depths have been translated into effective Earth sizes.
}
\end{figure}

Arguably, the most intriguing detections of {\hi} {\lalpha} line absorption are those on the Neptune-sized GJ 436 b, GJ 3470 b, and HAT-P-11 b \citep{kulowetal2014,bourrieretal2018,ben-jaffeletal2022}. All three planets show evidence for envelopes of {\hi} atoms out to many times their optical radii (Fig. \ref{lymanalpha_fig}) that mask large fractions of their host star disks. 
Given the strong measured signal brightness of the host stars, temporally-resolved spectroscopy is feasible, thereby enabling unique physical insight. 
For GJ 436, the most extensively investigated of these planets, 
the transmission spectrum is highly asymmetric, with absorption at blue (short) wavelengths typically stronger than at red (long) wavelengths. This has been interpreted as resulting from a long tail of {\hi} atoms that trails the planet at line-of-sight velocities of at least 120 km/s. 
The three-dimensional shape of this {\hi} envelope is presumably responsible for the specifics of the light curve (especially at blue wavelengths), which exhibits a sharp ingress but a very long egress \citep{lavieetal2017}. 
The atmospheric models that have attempted to reproduce the richness of the measured temporally-resolved spectra have mostly focused on the multi-dimensional dynamics of the planetary and stellar winds under the effects of stellar radiation and ram pressure, charge exchange between the stellar wind protons and the planetary neutrals, photoionization and the simultaneous gravitational pull from both the star and the planet \citep{khodachenkoetal2019,villarrealdangelo2021}. 
Although some published models reproduce reasonably well the main features of the measurements (possibly tuning the stellar XUV output and stellar wind properties), other features such as the egress light curve shape at red wavelengths have demonstrated to be difficulty to model, a difficulty that suggests that some physics in the star-planet interaction of the models is missing. 
The fact that the relative importances of stellar radiation and ram pressure, and charge exchange, is not entirely clear serves also as a warning that the star-planet interactions driving the strength of the {\hi} {\lalpha} line are not fully understood yet. More work is needed, in particular to understand the atom-scale details of charge exchange and the significance of diffuse {\lalpha} radiation -- effects that have so far been overlooked.
\\

There has been recent interest in using the {\hi} {\lalpha} line to explore sub-Neptune-sized exoplanets, 
thus complementing transmission spectroscopy observations at longer wavelengths and that probe the deeper atmospheres. 
The attempts to detect the absorption signature of this line on planets smaller than 3$R_{\oplus}$ include 
{\pimenc}, TRAPPIST-1 b \& c, HD 97658 b, Kepler-444 e, GJ 1132 b, 55 Cnc e, HD 63433 b \& c, GJ 9827 b, and TOI-776 b \& c.
Except for TOI-776 b \& c \citep{loydetal2025}, 
all these detection attempts 
have failed, which has been puzzling because strong signatures are expected if the planet atmospheres are hydrogen-dominated. This is indeed the case of {\pimenc} (2.1$R_{\oplus}$, 4.5$M_{\oplus}$), which orbits its Sun-like star every 6.3 days. 
A single-transit observation of this object \citep{garciamunozetal2020} with the HST/STIS instrument (Fig. \ref{lymanalpha_fig}) revealed no evidence of in-transit absorption even though the data are of high quality. Three-dimensional modelling of the system shows that the non-detection can only be explained by a stellar wind notably weaker than solar,  
an XUV higher than estimated from the available X-ray data or a combination of these two conditions \citep{shaikhislamovetal2020}. 
The non-detection might also be explained if the atmosphere is not hydrogen dominated, which would reduce the escape rate and accelerate the transition in the planetary wind from neutral into ionized hydrogen \citep{garciamunozetal2020}. 
The idea of a volatile-rich atmosphere is reinforced by the fact that the retention time of a hydrogen-dominated atmosphere on {\pimenc} is much shorter than its age ($\sim$5 Gyr). Observations of {\pimenc} at optical and infrared wavelengths, in particular with 
JWST, ARIEL \citep{tinettietal2018} or from the ground, have the potential to shed new light on the nature of this planet and in turn other sub-Neptunes.
\\

A few lines of the {\hi} Balmer series, H($n$=2)+$h\nu${$\rightarrow$}H($n${$\ge$}3), where $n$ stands for principal quantum number, have been detected on a few ultrahot Jupiters with equilibrium temperatures $>$2,000 K \citep{yanhenning2018,czeslaetal2022}, some of them orbiting early-type stars.  
According to models, the absorbing excited state H($n$=2) is populated by either {\lalpha} radiation (stellar plus diffuse) or by collisions of H($n$=1) with thermal electrons when the gas temperature is sufficiently high \citep{yanetal2021}. 
For the extreme KELT-9b ($T_{\rm{eq}}${$\sim$4,000 K}) models show that the excited state H($n$=2) (ionization potential of 3.4 eV) plays a key role in the ionization and heating of the atmosphere under the strong near-UV irradiation conditions of its A-type host star \citep{garciamunozschneider2019}. KELT-9b represents an exception, although possibly not the only one, in that the stellar near-UV  spectrum may be more important than at least as important as its XUV counterpart at governing the specifics of the atmospheric outflow.
Temporal variability during transit in some of the atmospheric lines detected on this planet have been tentatively ascribed to a flare \citep{cauleyetal2019}, challenging the common belief that A-type stars do not exhibit such events. Subsequent observations have not confirmed this possibility \citep{wyttenbachetal2020}.
Unlike the {\hi} {\lalpha} line at 1,216 {\AA}, those in the Balmer series are largely unaffected by the terrestrial atmosphere and can be observed from the ground at high spectral resolution. The lines are usually resolved over the core and wings, and the existence of multiple of these, each with a different strength as dictated by their specific oscillator strengths makes the Balmer series sensitive to a range of atmospheric altitudes. There are several hints that some of the {\hi} Balmer lines might be Doppler-shifted with respect to star rest frame (Fig. 25 of \citet{sanchez-lavegaetal2023}). Indeed, the significant width of the lines makes it difficult to extract any information concerning the preferential motion of the absorbing gas either towards the observer or the star.
\\

What is commonly called the {\cii} ``resonance line" at 1,334/1,335 {\AA} is actually a triplet. The component at 1,334 {\AA} is often masked by ISM absorption, with the two components at 1,335 {\AA} overlapping and being in practice undistinguishable. There is some evidence for the presence of this line in the atmospheres of HD 209458 b \citep{vidal-madjaretal2004}, {\pimenc} \citep{garciamunozetal2021} and HAT-P-11 b \citep{ben-jaffeletal2022}. Neutral carbon (IP=11.3 eV) becomes rapidly photoionized under the stellar irradiation conditions of close-in exoplanets. 
The photoionization lifetime of {\cii} (IP=24.4 eV) is however much longer, which ensures that once formed the {\cii} ion may travel far in the atmosphere and is largely unaffected by radiation, potentially building up a large absorbing column in the line-of-sight direction. This is essentially the idea used to rationalize the detections on both {\pimenc} and HAT-P-11 b of {\cii} with velocities of up to 70 km/s away from the star. As an ion, the {\cii} dynamics and therefore its corresponding line strength is sensitive to the  magnetic environment near the planet, as shown by combined hydrodynamic and Particle-In-Cell simulations \citep{ben-jaffeletal2022}.
\\

Transmission spectroscopy in the {\hei} triplet line at 1.08 $\mu$m has emerged in the last years as a valuable means of exploring the upper atmospheres of close-in exoplanets. The line is formed in the dipole-allowed transition He(2$^3S$)+$h \nu$(1.08 $\mu$m){$\rightarrow$}He(2$^3P$) that connects the metastable state He(2$^3S$) with the next excited state in the triplet manifold. The He(2$^3S$) population in the atmosphere is typically very small, but the large cross section of the transition at the line core and strong stellar background in the infrared make the line detectable at high resolution. The {\hei} triplet line is indeed an ideal target for observations from the ground with moderate-to-large telescopes equipped with high-resolution spectrographs. \\

The He(2$^3P$) state is populated by radiative recombination of the precursor {\heii} ion, as in {\hep}{+}{\eminus}{$\rightarrow$}He($i$)+$h\nu$ and subsequent de-excitation of the nascent excited states He($i$) \citep{oklopcichirata2018}. Large amounts of He(2$^3S$) are formed in the collision of non-thermal electrons with ground state He(1$^1S$), but this process is usually only the dominant one too deep in the atmosphere to be detectable in the transmission spectrum \citep{garciamunoz2025}.
The He(2$^3S$) loss is controlled by a variety of processes. Photoionization by stellar photons of wavelengths shorter than 2,600 {\AA} is potentially important 
\citep{oklopcic2019}, even if  the predictions based on it do not fully match observations \citep{guilluyetal2024}. 
Penning ionization, where the energy of the excited state He(2$^3S$) is used to eject an electron from a hydrogen atom, is also important \citep{garciamunoz2025}. 
\\

Interestingly, the {\hei} triplet line at 1.08 $\mu$m has been detected in a few Neptune- and sub-Neptune-sized planets \citep{guilluyetal2024}. The latter group includes GJ 3090 b \citep[$\sim$2.1$R_{\oplus}$, ][]{ahreretal2025}. The correct interpretation of the line presence is particularly important for these objects, which often lack any other signatures in their optical and infrared transmission spectra. On these planets, the removal of the precursor ion
{\hep} in collisions with {\htwo} at moderate temperatures acts as a efficient He(2$^3S$) sink \citep{garciamunozetal2025}. 
Omitting this process may bias any inferences on the He/H ratio in the planets' atmospheres.\\

Like the {\hi} {\lalpha} line, the {\hei} triplet line enables exploring the large-scale dynamics of the outflow. This is particularly important for planets such as the the ultrahot Jupiter HAT-P-32 b, which exhibit very extended (leading and trailing) tails traceable in the {\hei} triplet line \citep{czeslaetal2022,zhangetal2023}. 
For HAT-P-32 b and some other ultrahot Jupiters, the simultaneous detection of the {\hei} triplet line and some {\hi} Balmer lines provides complementary information to constrain the chemistry, dynamics and energetics of the gas. 
Unlike the {\hi} Balmer lines, which so far have only been detected at ultrahot Jupiters, the {\hei} triplet line has been identified on planets with very different equilibrium temperatures. 
The {\hi} Balmer lines, the {\hei} triplet line is also relatively narrow, which allows to determine its relative position with respect to the stellar rest frame. Many of the {\hei} triplet line detections exhibit a net blue shift that suggests a day-to-night net motion of the He(2$^3S$) atoms. According to models \citep{schreyeretal2024}, the net offset may be sensitive to the strength of the near-planet magnetic field. If true, the measured offset would be revealing the planet's magnetic field. 
For some planets such as the warm-Neptune GJ 3470 b, the strength of the {\hei} triplet line seems to vary significantly over time 
\citep{guilluyetal2024,massonetal2024}. Model calculations show that the varying line strength is consistent with a possible rotational or cycle modulation in the stellar output from the host star \citep{garciamunozetal2025}. 
The hypothesis should ideally be tested with simultaneous observations of the {\hei} triplet line and the host star's X-ray output. If proven correct, the available and future observations of the line might be used to monitor the stellar high-energy output.

While transmission spectroscopy provides a powerful probe of exoplanetary atmospheres and their coupling to the stellar environment, as discussed above, its interpretation is intrinsically limited by the fact that the observed signal is a convolution of planetary atmospheric absorption and the heterogeneous stellar photosphere.
Stellar photospheric heterogeneities, such as starspots and faculae, introduce wavelength-dependent biases in transmission spectra through the so-called transit light source effect (TLSE, \citet{Rackham2018}), which can mimic or obscure atmospheric features and bias the inferred planet-to-star radius ratio. Recent studies have shown that simplified contamination models based solely on disk-averaged filling factors and spectral contrasts may fail to accurately reproduce the observed signal, particularly in the optical where limb darkening and transit geometry play a dominant role \citep{Sumida2026}. Self-consistent, pixel-resolved modeling demonstrates that these effects can lead to discrepancies of up to several hundred ppm, comparable to or larger than typical atmospheric signals, especially for active stars and non-equatorial transits. Moreover, attributing observed spectral slopes solely to stellar contamination may require unrealistically large and hot active regions, indicating that a combination of stellar and planetary contributions is often necessary to explain the data. These results highlight that reliable atmospheric characterization demands geometry-aware models of stellar heterogeneity that explicitly account for the spatial distribution of active regions and limb-darkening effects.

\vspace{2.0cm}

\section*{SPI detection in timeseries}

Ultimately, the detection of SPI signatures relies on their unambiguous identification in time series measurements. Signal searches are often conducted by attempting to identify periodic or quasi-periodic modulations at characteristic timescales of the system, such as the planetary orbital period, the stellar rotation period, or the synodic period between the two.

These time series are typically analysed using generalized Lomb–Scargle (GLS) periodograms \citep{Lomb1976, Scargle1982, Zechmeister2009}, which are well suited to the unevenly sampled data commonly acquired for planetary searches and characterization. In particular, a recurring modulation phased with the planetary orbit is often regarded as one of the most compelling signatures of SPI, and several methodological advances toward this goal have been developed in recent years.

\subsubsection*{Characteristic periods and pre-whitening}

A fundamental preliminary step in any time-series search for SPI is the identification and removal of signals unrelated to the planet. The dominant timescale and mechanism of stellar variability is stellar rotation: active regions rotating in and out of view modulate the emission at the stellar rotation period, $P_\mathrm{rot}$, and its harmonics. SPI-induced activity, by contrast, is not anchored to the stellar surface; instead, it is synchronized with the planet’s orbital motion and is therefore expected to be modulated at a period related to the planetary orbital period, $P_\mathrm{orb}$ \citep[e.g.][]{Shkolnik2003,Shkolnik2005}.

However, the SPI signal is not necessarily expected to appear exactly at $P_\mathrm{orb}$. Considering the case of SPI measured through a chromospheric index, the interaction may may preferentially onto localized chromospheric regions fixed in the stellar reference frame, while the observable modulation occurs at the synodic period of the system \citep{fares2010,Lanza2012}:

\begin{equation}
  P_\mathrm{syn} = \left(\frac{1}{P_\mathrm{orb}} - \frac{1}{P_\mathrm{rot}}\right)^{-1},
\end{equation}

Depending on the geometry of the interaction, planets on polar or retrograde orbits may also exhibit a complementary frequency, the so-called anti-synodic period, $P_\mathrm{a-syn} = (P_\mathrm{orb}^{-1} + P_\mathrm{rot}^{-1})^{-1}$ \citep{Revilla2026}. A search restricted to $P_\mathrm{orb}$ alone therefore risks missing the actual SPI signal, and periodogram analyses should generally explore the vicinity of all four characteristic periods.

\cite{FischerSaur2019} proposed additional periods at which SPI signatures might manifest and that should therefore be considered in such searches. Depending on the system geometry and viewing angle, a signal may be observable at $P_{\rm syn}/2$. Moreover, in systems hosting multiple planets within the Alfvén surface, each planet may generate its own Alfvén wing. Interactions between these wings could then produce a signal with a characteristic period corresponding to the synodic period between the orbital periods of the planets involved.

The expected periodicity of the SPI signal also depends on the underlying interaction mechanism.
Magnetic SPI is often expected to produce variability at the synodic period, or at related orbital frequencies, since the interaction is governed by the relative motion between the planet and the stellar magnetic field. 
However, magnetic SPI signatures may also appear at harmonics of the orbital period, including $P_{\rm orb}/2$, for instance when the interaction geometry produces two maxima per orbit, as observed in radio studies of the Proxima Centauri system \citep[e.g.][]{PerezTorres2021}, in analogy with the Jupiter--Io interaction.
Tidal SPI, on the other hand, can naturally generate signals at harmonics of the orbital period, particularly at $P_{\rm orb}/2$, owing to the formation of two tidal bulges on opposite hemispheres of the star \citep[e.g.][] {Cuntz2000}. 
Consequently, searches for SPI signatures should account not only for the orbital and synodic periods, but also for their relevant harmonics.

In extreme close-in systems, tidal dissipation and magnetic torques may also lead to secular orbital evolution, causing the orbital period itself to vary with time. In such cases, SPI signatures are not expected to remain stable over long temporal baselines, but will instead show a gradual phase drift associated with orbital decay \citep[e.g.][]{Yee2020,Strugarek2017}. This effect can broaden or shift peaks in classical periodograms and may require searches based on time-dependent ephemerides rather than a constant orbital period. Such behaviour is particularly relevant for ultra-short-period planets and systems undergoing strong tidal or magnetic dissipation.

\subsubsection*{Dataset splitting and rolling periodograms}

One critical complication that distinguishes SPI searches from conventional period searches is the intermittent, or ``on/off'', nature of the interaction itself. Observational evidence shows that SPI signals can appear and disappear across different epochs, possibly as a consequence of changes in the stellar magnetic topology over the activity cycle or variations in the coupling efficiency as the star’s large-scale magnetic field evolves \citep[e.g.][]{Shkolnik2008,Lanza2012,Revilla2026}. When the interaction is active during only a subset of the observations, combining all available data into a single set effectively dilutes the signal, thereby reducing its power in the periodogram.

One way to address this problem is to divide the full dataset into shorter, contiguous subsets, each spanning a single observing season or a well-defined temporal epoch, and analyse them independently \citep{Shkolnik2003,Revilla2026}. This approach mitigates signal dilution at the expense of a shorter temporal baseline and requires that each subset contain enough observations to resolve the periods of interest. The choice of subsets or timescales should also be guided by physical considerations and external information, as it directly impacts the detectability of potential signals.

An alternative strategy to factor in the evolving nature of SPI signals is to use rolling periodograms \citep{Herbort2018,Schofer2021}. Rather than dividing the data into a fixed set of non-overlapping blocks, the rolling periodogram slides a window of a fixed size $m$ of observations along the time series, computing a GLS periodogram for each window position. The collection of periodograms is then displayed as a two-dimensional map of power as a function of frequency and time, allowing to visualize the emergence, persistence, and decay of periodic signals across the observation record. The window length $m$ represents a trade-off: too short a window fails to resolve closely spaced periods, while too long a window averages over transient behaviour. Rolling periodograms are particularly well suited to the study of SPI because they not only reveal whether a signal is present, but also identify the epochs during which the interaction is strongest, thereby guiding the selection of the most informative sub-datasets for detailed analysis.

\subsubsection*{False alarm probability and bootstrap significance}

Assessing the statistical significance of a periodogram peak is therefore central to any detection claim. The standard metric is the False Alarm Probability (FAP), defined as the probability that a signal with the observed power, or greater, could arise purely from noise. Analytical estimates of the FAP are readily available for Lomb–Scargle-type periodograms \citep{Baluev2008,Zechmeister2009}, but they rely on assumptions of Gaussian white noise and weakly correlated sampling that are rarely satisfied in practice.
In SPI searches, additional complications arise from red noise associated with evolving stellar activity, irregular sampling, and the fact that the frequency of interest is often known a priori from the system ephemeris. In this latter case, the relevant question is not the probability of finding a strong peak anywhere in the spectrum (the global FAP), but rather the probability of finding such a peak at a specific, predetermined frequency (the local FAP). Since the number of independent frequencies being tested is effectively unity, the local FAP can be orders of magnitude smaller than the global FAP, making it a substantially more informative quantity for assessing the significance of candidate SPI signals.

A robust approach for estimating the local FAP is bootstrap randomization combined with a frequency-windowing technique \citep{Hatzes2019}. The procedure consists of randomly shuffling the observed data values while keeping the observation timestamps fixed, thereby preserving the spectral window and the overall statistical properties of the original dataset. The GLS periodogram is then recomputed for a large number of realizations ($\sim 10^5$--$10^6$), and the fraction of cases in which the power within a narrow window centred on the frequency of interest exceeds the observed power provides the bootstrap FAP.
Repeating the procedure with progressively narrower windows and extrapolating to zero width yields an empirical estimate of the probability that random noise alone could reproduce the observed signal at the exact frequency under consideration, rather than anywhere within a broader frequency interval. This approach is less sensitive to the assumed noise model than analytical estimates and is generally more robust against departures from ideal white-noise assumptions, including non-Gaussian statistics and weakly correlated residuals.

\vspace{1.2cm}

Taken together, these considerations illustrate that the detection of SPI signatures requires methodologies capable of accounting for complex stellar variability, irregular sampling, evolving signals, and prior knowledge of the characteristic interaction timescales. Robust identification therefore relies not on a single significant signal, but rather on the detection of a coherent physical pattern across time, consistent with the expected behaviour of star--planet interactions and, ideally, across multiple diagnostics. In this context, an ideal approach would involve a formalism capable of distinguishing between SPI-affected and SPI-unaffected measurements, thereby isolating the contribution of the interaction from the intrinsic variability of the host star.

\vspace{2.0cm}

\section*{Concluding remarks}

The observational study of star–planet interactions (SPI) remains challenging, primarily because these signatures are often subtle, intermittent, and difficult to disentangle from intrinsic stellar variability. As discussed throughout this chapter, a wide range of diagnostics—including radial velocities, precision photometry, chromospheric activity indicators, radio observations, and transmission spectroscopy—provide complementary probes of SPI across different atmospheric layers and physical regimes. Robust identification requires consistent evidence across multiple observables, timescales, and analysis methods.

A central theme emerging from current studies is that stellar variability is not merely a source of noise, but a fundamental limitation that must be modeled carefully. Advances in time-series analysis, correlated-noise modeling, and statistical validation have significantly improved our ability to assess the significance of potential SPI signals. At the same time, geometry, stellar magnetic topology, and observational biases play a critical role in shaping the detectability and interpretation of these signals.

Radio observations, in particular, offer a unique and potentially direct probe of magnetic SPI through coherent emission mechanisms, with the ability to constrain planetary magnetic fields and magnetospheric interactions via polarization, spectral, and temporal diagnostics.
Even non-detections in radio can place meaningful constraints on stellar wind properties and interaction regimes.

Transmission spectroscopy and chromospheric diagnostics provide particularly valuable insights into the coupling between stellar radiation, winds, and planetary atmospheres, while photometric techniques such as spot occultations offer spatially resolved information on stellar surfaces and potential orbit-locked activity. Nevertheless, stellar heterogeneity, especially starspots and faculae, remains a major source of contamination, affecting both SPI diagnostics and atmospheric characterization.

Looking ahead, progress in this field will depend on combining multi-wavelength observations with physically motivated models that account for stellar magnetic structure and temporal variability. The synergy between current and upcoming facilities, including high-resolution spectrographs and space-based missions such as PLATO and HWO, will be essential to move from tentative detections toward a more systematic and physically grounded understanding of star–planet interactions.

\section*{Acknowledgements}

 Pedro Figueira thanks Pedro Amado for insightful comments on an early version of the manuscript.
 
 PF, LPM, MPT, and DR acknowledge financial support from the Severo Ochoa grant CEX2021-001131-S funded by MCIN/AEI/10.13039/501100011033. 
 
 PF is funded by the European Union (ERC, THIRSTEE, 101164189). Views and opinions expressed are however those of the author(s) only and do not necessarily reflect those of the European Union or the European Research Council. Neither the European Union nor the granting authority can be held responsible for them.

 LPM and MPT also acknowledge financial support from the Spanish grant PID2023-147883NB-C21, funded by MCIU/AEI/ 10.13039/501100011033, as well as support through ERDF/EU.

 PC acknowledges the support of the Department of Atomic Energy, Government of India.
 
 RF acknowledges support from the United Arab Emirates University (UAEU) UPAR grant number G00005451. 

\vspace{2.0cm}

\bibliographystyle{plainnat}

\bibliography{bib_files/RV,bib_files/photometry,bib_files/chromospheric_indexes,bib_files/radio,bib_files/spectroscopy,bib_files/timeseries}

\end{document}